\documentclass[a4paper,onecolumn,9pt]{article}
\usepackage{extsizes}
\usepackage[super,sort&compress,comma]{natbib} 
\usepackage[version=3]{mhchem}
\usepackage[left=1.5cm, right=1.5cm, top=1.785cm, bottom=2.0cm]{geometry}
\usepackage{balance}
\usepackage{times,mathptmx}
\usepackage{sectsty}
\usepackage{graphicx} 
\usepackage{lastpage}
\usepackage[format=plain,justification=justified,singlelinecheck=false,font={stretch=1.125,small,sf},labelfont=bf,labelsep=space]{caption}
\usepackage{float}
\usepackage{fancyhdr}
\usepackage{fnpos}
\usepackage[english]{babel}
\addto{\captionsenglish}{%
  
}
\usepackage{array}
\usepackage{droidsans}
\usepackage{charter}
\usepackage[T1]{fontenc}
\usepackage[usenames,dvipsnames]{xcolor}
\usepackage{setspace}
\usepackage[compact]{titlesec}
\usepackage{hyperref}

\usepackage{epstopdf}

\begin{document}

\LARGE{\textbf{Titania-based spherical Mie resonators elaborated by high-throughput aerosol spray;\\ single object investigation}} \\
\\
\large\textit{{
Simona Checcucci\textit{$^{a, b, c}$},
Thomas Bottein\textit{$^{a}$},
Jean-Benoit Claude\textit{$^{a}$},
Thomas Wood{$^{a}$}$^{\dag}$,
Magali Putero\textit{$^{a}$},
Luc Favre\textit{$^{a}$},
Massimo Gurioli\textit{$^{a}$},
Marco Abbarchi\textit{$^{a}$}$^{\ast}$,
David Grosso\textit{$^{a}$}$^{\ast\ast}$
}} \\
\\
\textit{$^{a}$NOVA Team, CNRS, Aix-Marseille Universit\'{e}, Centrale Marseille, IM2NP, UMR 7334, Campus de St. J\'{e}r\^{o}me, 13397 Marseille, France.}\\
\textit{$^{b}$European Laboratory for Nonlinear Spectroscopy (LENS), 50019 Sesto Fiorentino, Italy.}\\
\textit{$^{c}$Dipartimento di Fisica ed Astronomia, Universit\'{a} degli Studi di Firenze, 50019 Sesto Fiorentino, Italy.}\\
\textit{$^{\dag}$Present address: INL, Ecole Centrale de Lyon, 36 Avenue Guy de Collongue,
69134 \'{E}cully, France.}\\
\textit{$^{\ast}$Corresponding author: marco.abbarchi@im2np.fr}\\
\textit{$^{\ast\ast}$Corresponding author: david.grosso@im2np.fr}
\\
\noindent\normalsize{
In the framework of photonics with all-dielectric nano-antennas, sub-micrometric spheres can be exploited for a plethora of applications including vanishing back-scattering, enhanced directivity of a light emitter, beam steering and large Purcell factors. Here we show the potential of a high-throughput fabrication method based on aerosol-spray, to form quasi-perfect sub-micrometric spheres of poly-crystalline TiO$_2$.
Spectroscopic investigation of light scattering from individual particles revealed sharp resonances in agreement with Mie theory, neat structural colours and a high directivity. Owing to the high permittivity and lossless material in use, this method opens the way towards the implementation of isotropic meta-materials and forward-directional sources with magnetic responses at visible and near-UV frequencies, not accessible with conventional Si- and Ge-based Mie resonators.
} \\
\\
%
%
%
%
%
%
%

\renewcommand*\rmdefault{bch}\normalfont\upshape
\rmfamily
\section*{}
\vspace{-1cm}



%


\section{Introduction}
Dielectric Mie resonators  represent a young and promising research topic, with huge potentials in photonics, optics and derived applications\cite{Brongersma2014,Kuznetsov2016,jahani2016}. Based on sub-micrometric particles that are capable of light manipulation at visible and near-infrared frequencies, these systems offer convenient and ultra-thin alternatives to replace complex and bulky optical elements. This unique ability is due to strong modifications of the local density of optical states occurring in sub-micrometric objects made of materials featuring high dielectric constant and sufficiently small absorption losses.  

Most studies over the last years have mainly addressed silicon-\cite{FEN2008,Vynck2009,Evlyukhin2010,Garcia2011,Van2013,Staude2017,SHE2018} and germanium-based\cite{Gomez2011,Zhu2017,Wood2017,BOU2018} Mie resonators, demonstrating that they could outperform their metallic counterpart supporting localized plasmonic resonances. However, the large absorption of group IV semiconductor compounds at short wavelengths induce strong optical losses, limiting their potential applicability as efficient devices especially at blue and near-UV frequencies\cite{PHI1959,PHI1960} (e.g. at 450 nm: n$_{Si}$ = 4.5, k$_{Si}$ = 0.13; n$_{Ge}$ $=$ 4.0, k$_{Ge}$ $=$ 2.24). Furthermore, with a few exception based on colloids\cite{FEN2008,Garin2014,Shi2012} and solid state dewetting\cite{Wood2017,Abbarchi2014,Naffouti2016,Naffouti2016c,Naffouti2016b,Naffouti2017}, typical nano-fabrication methods of Si(Ge)-based Mie resonators rely on top-down technologies that are not easy to scale-up at affordable prices.
 
TiO$_2$-based optical devices are an interesting alternative to Si, since Titania has a relatively high refractive index and is fully transparent up to UV frequencies\cite{Kim1996,Siefke2016} (e.g.: at 450 nm: n$_{TiO_2}$ = 2.55, k$_{TiO_2}$ = 1.2$\cdot$10$^{-5}$; at 370 nm: n$_{TiO_2}$ = 2.83, k$_{TiO_2}$ = 1$\cdot$10$^{-3}$) rendering it, for instance, a strategic material to manipulate the light emitted by conventional GaN-based blue LEDs ($\sim$450 nm). TiO$_2$ can be  prepared by high-throughput chemical processes, which is a prerequisite for applications requiring large surface systems. It also has many other advantages over Si and metals, that are its high chemical, mechanical and thermal stability, non-toxicity, and relative natural abundance.

To date, several groups have studied the properties of Titania particles as dielectric resonators prepared using conventional top-down microfabrication technologies \cite{Khorasaninejad2016,Gutruf,Devlin2016,Sun2017,SUN2018} or soft-nano imprint lithography\cite{Bottein,Bottein2018}. They all confirmed that electromagnetic resonances could be generated within these metal oxide objects. However, the limited exploitation of this material is mainly due to the difficulty in applying conventional top-down fabrication methods to TiO$_2$.  Additionally, such approaches do not allow the preparation of spherical resonators\cite{TZA2018}, which may be interesting for many applications with effective metamaterials, such as beam steering and  back-scattering-free optics\cite{Gomez2011,Krasnok2011,Fu2013,CAM2015,Zhang2015,Yan2015,Liberal2015,Albella2015,Tribelsky2015,Wozniak2015,Shibanuma2016,OLL2018,KOM2018}, enhanced light extraction\cite{Krasnok2015,VAS2018}, resonant transmission\cite{Savelev2015}, manipulation with optical tweezers\cite{Andres2016}, light detectors\cite{Garin2014}, directional Fano resonances\cite{Yan2015b,Tribelsky2016} and much more. 
 
Here, we focus on the investigation of the scattering properties of individual, dense, spherical TiO$_2$ particles of different sizes. The particles were prepared by a high throughput aerosol generation method on silicon and on silica substrates. Their structural properties were characterized by scanning electron microscopy and X-ray diffraction, whereas their scattering properties were assessed using dark field microscopy and spectroscopy. The optical resonances observed in experiments are interpreted on the basis of analytical solutions of Mie scattering from a sphere in vacuum and by finite difference-time domain simulations in vacuum, on silica and on silicon, in order to address the role of the resonance coupling with the underlying substrate. We demonstrate that spherical Titania particles are efficient scattering objects and can be tuned in size to scatter in a relatively broad range of frequencies (from blue to near-infrared). Sharp and bright resonances can be observed for particles deposited on a low-refractive index substrate (e.g. silica) and on high-reflective index substrate (e.g. silicon) accounting for the robustness of this approach.

%

\section{Results and discussion}

\begin{figure}[t!]
\centering
\includegraphics[width=0.8\textwidth]{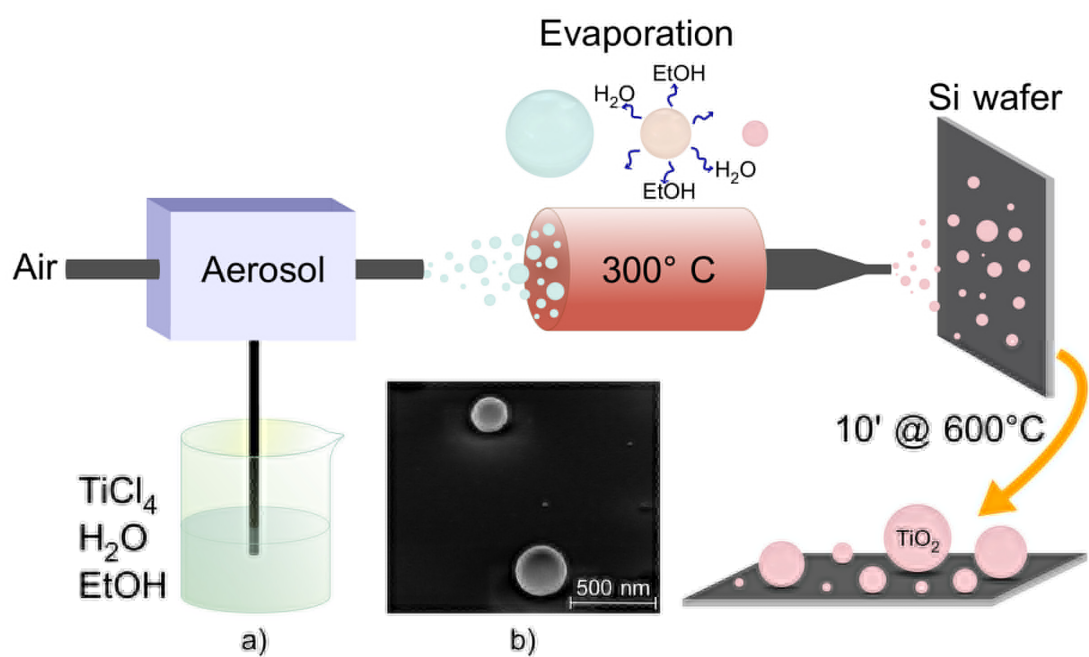}
\caption{(a) Schematic representation of the process used to prepare dense TiO$_2$ Anatase particles on a Si wafer. (b) SEM image of two distinct spheres deposited on Si taken at 55 degrees tilting angle.}
\label{fgr:chem}
\end{figure}

Supported spherical TiO$_{2}$ particles were prepared through atomization of a sol-gel solution and deposition by direct impaction onto the surface of a silicon wafer (at 450 nm n$_{Si}$ $=$ 4.6), or a microscope glass slide (at 450 nm n$_{SiO_{2}}$ $=$ 1.5), before calcination at 600$^{\circ}$C  and 500$^{\circ}$C, respectively, to ensure densification. The particles deposited on a Si wafer where then characterized by high-resolution scanning electron microscopy (SEM) and x-ray diffraction (GIXRD). Spectroscopic measurements of light scattering were performed on particles formed on silicon and silica. Finally, in order to assure a direct comparison of the same object on each substrate, some particles were transferred from the Si surface to PDMS (PolyDiMethylSiloxane, PDMS; at 450 nm n$_{PDMS}$ $=$ 1.42). The elaboration protocol is illustrated in Fig.\ref{fgr:chem} whereas the associated details are provided in the Experimental Section.

\begin{figure}[t!]
\centering
\includegraphics[width=1\textwidth]{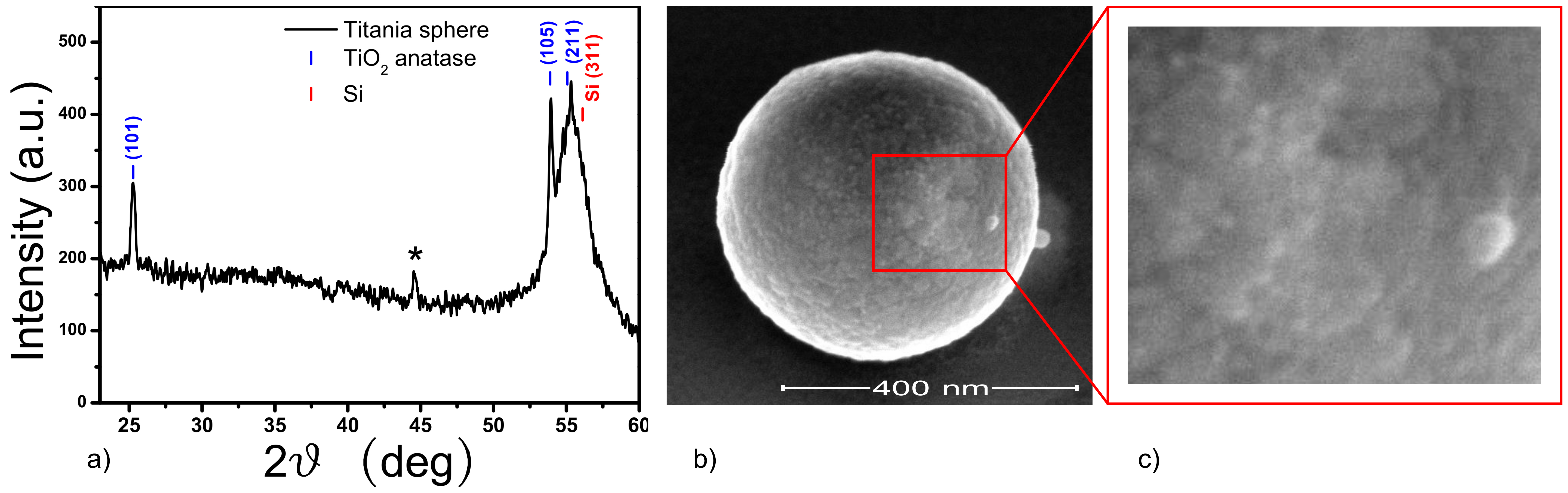}
\caption{(a)  GIXRD pattern of Titania spheres deposited onto a Si (001) substrate. The diffraction pattern is indexed with PDF \#00-021-1272 for the TiO$_2$ Anatase phase. The peak indexed with a star likely corresponds to a parasite peak due to unknown impurities from the process. (b-c) High-resolution SEM image at high magnification of a typical crystalline Titania sphere.}
\label{fgr:sphere}
\end{figure}

\begin{figure*}[t!]
\includegraphics[width=1\textwidth]{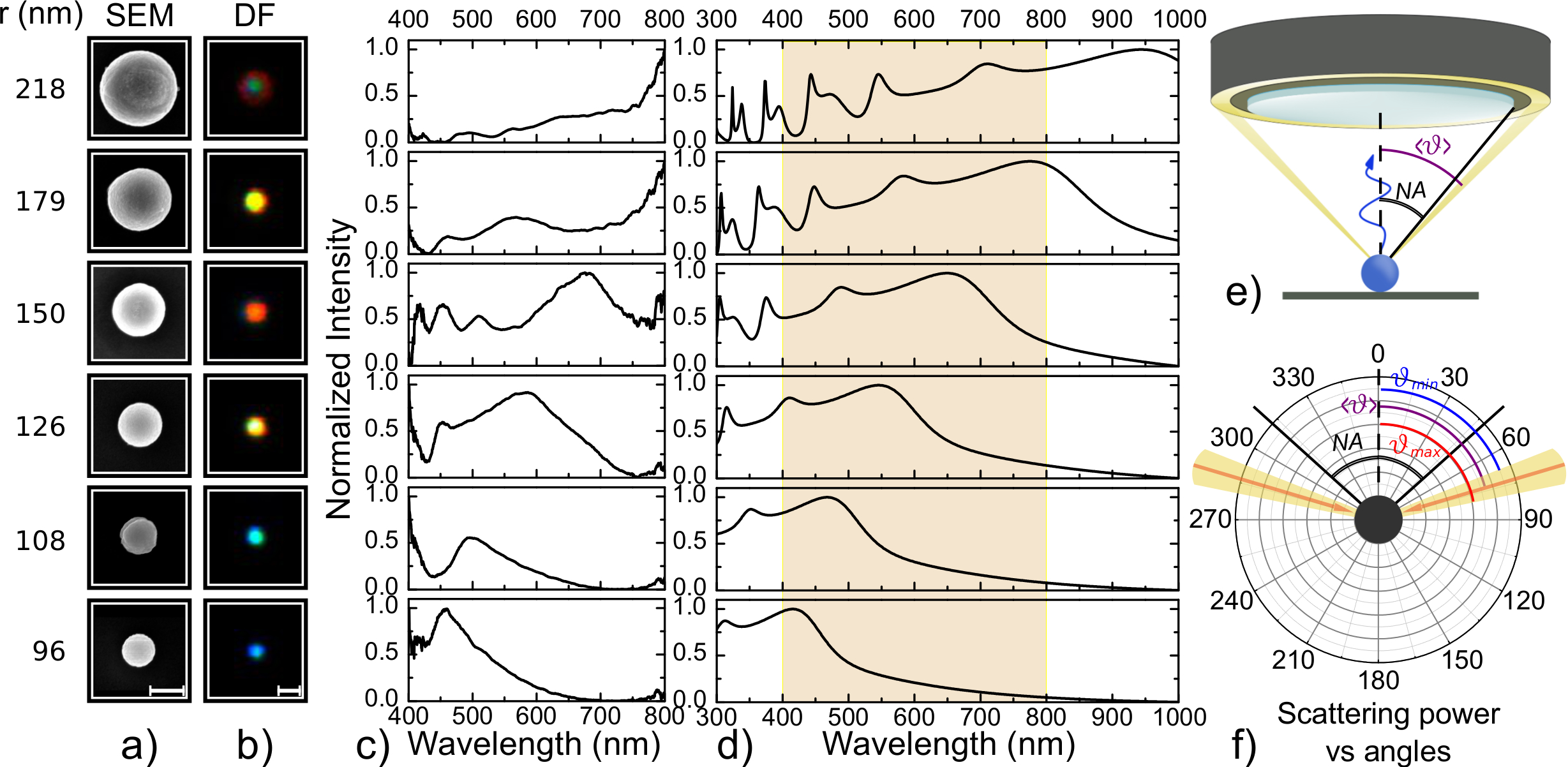}
\caption{(a) From bottom to top: high-resolution SEM images of TiO$_2$ spheres on a Si substrate with increasing radius size. The scale bar corresponds to 200 nm. (b) From bottom to top: dark-field optical microscope images of the TiO$_2$ spheres shown in (a).  The scale bar corresponds to 500 nm. (c) From bottom to top: scattering spectra produced by the TiO$_2$ spheres shown in (a) and (b). (d) Analytical models of the scattering spectra of TiO$_2$ spheres in vacuum for radii corresponding to those shown in (a), (b) and (c). The shaded area highlights the spectral range accessible in experiments. (e) Scheme of the excitation/collection geometry used in experiments. NA is the numerical aperture of the objective lens spanning over an angle of $\pm$48.6 degrees; <$\vartheta$> is the average illumination angle determined by the condenser of the microscope (see (f) for details). (f) Polar diagram highlighting the excitation and collection geometry used in experiments and theory (the TiO$_2$ sphere is represented as the black circle at the centre of the polar diagram). The theoretical excitation angle was chosen to be the average between the maximum and minimum experimental excitation angles (respectively $\vartheta_{max}$ and $\vartheta_{min}$). The collection angle within the NA of the objective lens is also highlighted.}
\label{fgr:exp}
\end{figure*}

SEM investigation revealed that the Titania particles have a spherical shape (as evidenced in Fig.\ref{fgr:chem}(b), \ref{fgr:sphere}(a) and \ref{fgr:exp}(a). This  is ascribed to the fact that aerosol generation atomizes liquid, spherical, micro-droplets that keep their morphology upon evaporation of water and ethanol within the furnace. This fast evaporation also initiates the poly-condensation of the Titania precursors. As a consequence, a stiff shell is formed on the spheres skin, preventing their collapse upon impaction on the receiving surface and keeping a small contact point with the underlying substrate.

The GIXRD pattern measured on a substrate with densely packed spheres reveals that they have crystallized under Anatase structure (Fig.\ref{fgr:sphere}(b)).  With the GIXRD geometry used for this analysis  the Si(311) Bragg reflection is also detectable around 2$\vartheta$ = 55 degrees. This is ascribed to the X-Ray beam average divergence and to the huge substrate signal. From the fitting of the Anatase (101) Bragg reflection, the broadening of the peak corrected from the beam footprint was calculated to be H$_c$ = 0.29 degrees, giving an average crystallite size of 27 $\pm$ 3 nm. More details are provided in the the Experimental Section.
 
These spheres are thus composed of densely packed particles building blocks with a size of the order of tens of nanometers as also confirmed by the SEM images (Fig. \ref{fgr:sphere}(b) ad (c)). Thus, the surface roughness SR is estimated to be much lower than the wavelength in the visible range (SR $\sim$ $\lambda$/50), so that the sphere surface can be considered as smooth, with negligible influence on the Mie scattering features. 

The optical constants of the material composing the particles cannot be measured directly on the spheres. Thus, they were deduced from ellipsometry investigation on a thin, plain  film. The ellipsometric $\phi$ and $\Delta$ curves have been  fitted with a Cauchy dispersion with k = 0 and n($\lambda$) $=$ A + B/$\lambda^2$  (with A $=$ 2.007, B $=$ 0.058), which gives n $=$ 2.293 at 450 nm.

Due to such relatively high index of refraction combined with a high transparency window for wavelength longer than $\sim$350 nm and inter-band transitions outside the visible spectral range, TiO$_{2}$ spheres are expected to show Mie resonances in the visible range for dimensions above 200 nm in diameter\cite{Garcia2011,Zhang2015}. Therefore, spheres featuring diameters ranging from about 200 to 500 nm were first spotted by SEM on a Si wafer, then imaged in dark field optical microscopy. Thus, the corresponding scattering spectra were collected via confocal microscopy.

SEM images of TiO$_2$ spheres on Si with their measured radius (r), corresponding DF images, as well as experimental and theoretical scattering spectra are displayed in Fig.\ref{fgr:exp}(a)-(d). The excitation/collection geometry used in experiments and theory is shown in Fig.\ref{fgr:exp}(e) and (f). This analysis shows that, when increasing the size of the particle, the wavelength of the corresponding fundamental resonance (the broader and more intense peak in the spectrum) red-shifts accordingly. This behavior is also reflected on the optical appearance of the spheres (Fig.\ref{fgr:exp}(b)), where one can observe a marked change of colour corresponding to a shift of the fundamental resonance. 

\begin{figure}[t!]
\centering
\includegraphics[width=0.6\textwidth]{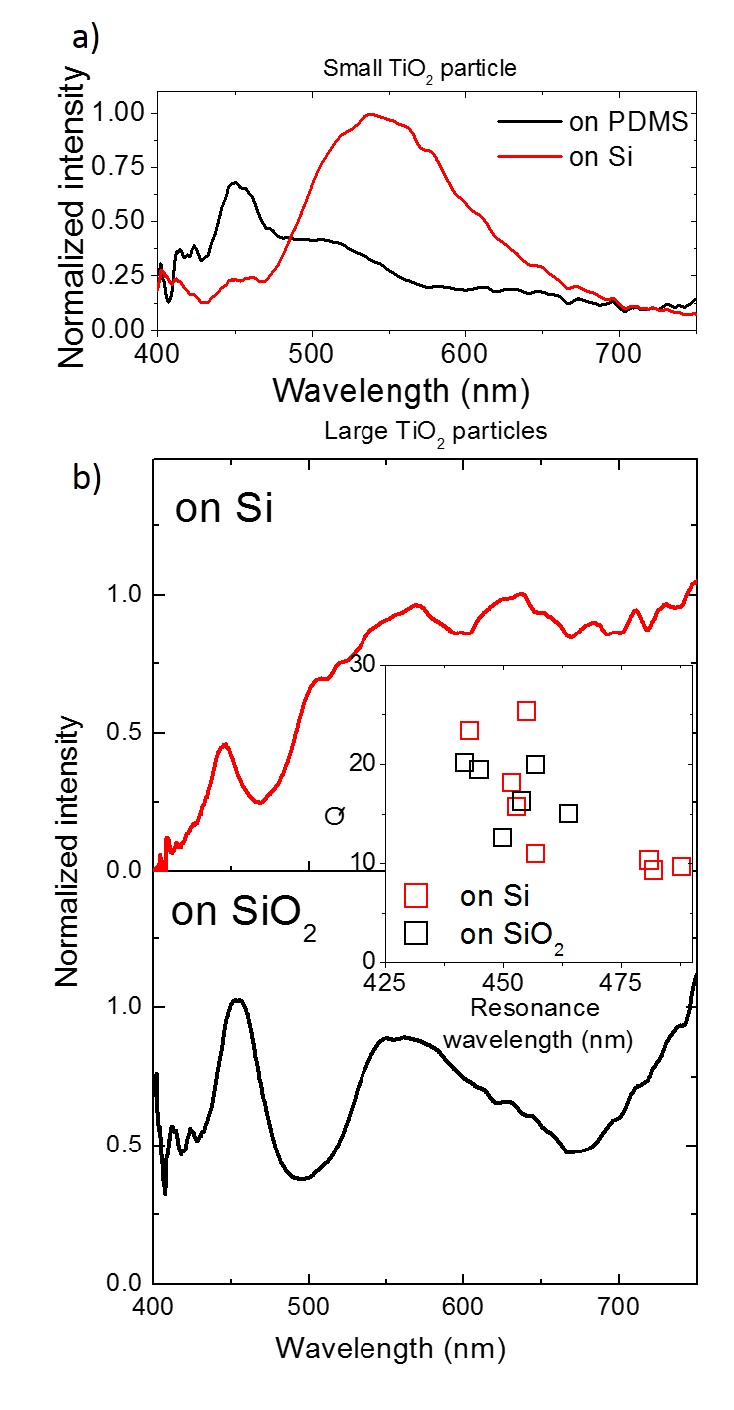}
\caption{a) Dark-field scattering spectra of the same small particle on a silicon substrate before transfer (red line) and after transfer on a PDMS slice (black line).  b) Top panel: dark-field scattering spectrum of a large particle on a silicon substrate. Bottom panel: dark-field scattering spectrum of a large particle on a silica substrate. The central inset shows the measured Q factors of resonances below 500 nm found in large spheres on Si (red squares) and on SiO$_{2}$ (black squares) as a function of the resonance central wavelength.}
\label{fgr:Si_vs_PDMS}
\end{figure}

Multiple and sharper peaks emerge in the spectra at shorter wavelength with respect to the main resonance (starting from r $=$ 126 nm), accounting for the onset of higher order modes  within the Titania spheres. Finally, for larger particles (r $>$ 180 nm) the fundamental mode shifts to near-infrared frequencies (non-accessible with our spectroscopic setup) and the higher order modes quench. However, for the sake of thoroughness, we mention that from a systematic investigation of particles on the same Si wafer, it emerges that large particles often exhibit sharp resonances below 500 nm as discussed later (Fig.\ref{fgr:Si_vs_PDMS} b)). 

This picture corresponds fairly well to that of a dielectric sphere scattering the impinging light in the far field as magnetic or electric multi-polar modes: the lowest order resonance is ascribed to the magnetic dipole, expected at a wavelength of about $\lambda_{MD}$ $\approx$ 2r$\cdot$n\cite{Evlyukhin2010,Garcia2011}: as an example, for a sphere featuring a radius of 96 nm and n $=$ 2.3, $\lambda _{MD}$ is expected at about 440 nm, not far from the peak measured in the experimental case at about 460 nm (bottom panel of Fig.\ref{fgr:exp}(c)). At shorter wavelength with respect to $\lambda _{MD}$ is found the electric dipolar mode, followed by higher order modes such as magnetic and electric quadrupols.  

In support of these experimental results, we calculated the analytical scattering spectra (Fig.\ref{fgr:exp}(d)) taking into account the excitation/collection geometry (Fig. \ref{fgr:exp}(e) and (f)). At this stage, the spheres are assumed to be in vacuum. Thus, the presence of the Si substrate is completely neglected. In the investigated range of spheres size, the fundamental resonance shifts from blue to near-infrared frequencies, together with the onset of several other peaks of increasingly higher Q-factor (defined here as the central wavelength of the resonance divided by its FWHM). This behavior is expected when moving from the Mie scattering approximation towards the whispering-gallery resonator picture \cite{Coenen} and, despite the oversimplified model, corresponds fairly well to the experimental findings at least for particles radius below 150 nm. 

These results account for the possibility to exploit our particles for efficient light manipulation from near-UV up to near-infrared frequencies, confirming the relevance of these spherical Titania-based resonators so far limited to theraherz frequencies\cite{Nvemec2012,Navarro2013}. The resonances found in experiments fairly agree with a very simple, analytical model for isolated perfect spheres predicting a red-shift of the resonances and the appearance of multiple peaks at shorter wavelength featuring higher Q factors. The discrepancies between experiments and theory are ascribed to small deviations from a perfect spherical shape and, above all, to the presence of the partially reflecting and high-refractive index, silicon substrate as discussed in more detail later\cite{Spinelli,Van2013,Markovich}.

\begin{figure}[t!]	
\centering
\includegraphics[width=0.4\textwidth]{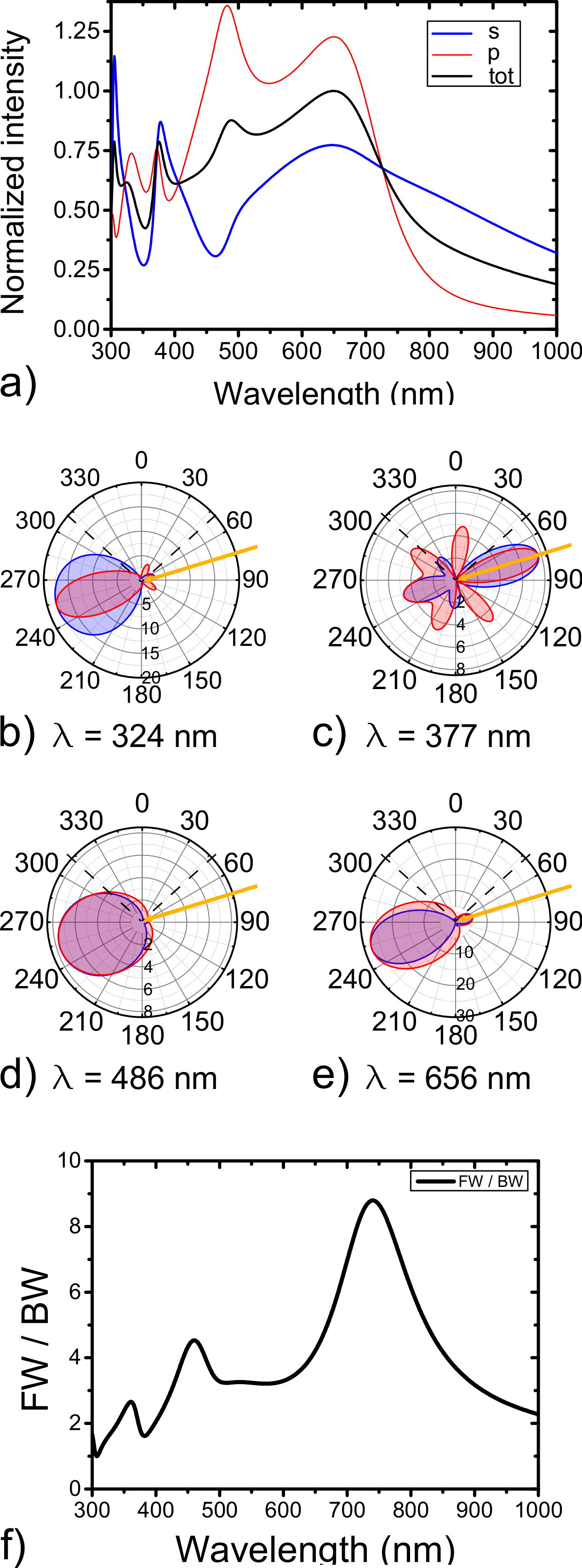}
\caption{(a) Analytical calculation of the scattering spectrum as a function of wavelength for a TiO$_2$ sphere having radius r = 150 nm in vacuum  (see also the corresponding panels in Fig.\ref{fgr:exp} (a)-(d). $s$ and $p$ polarizations are represented respectively as red and blue lines whereas their average is in black. The detection conditions are the same as those of the experimental case, as shown in Fig.\ref{fgr:exp} (e), (f). (b)-(e) Polar plots of the far-field scattering intensity for $s$ (red) and $p$ (blue) polarizations at the wavelengths corresponding to the main peaks seen in (a). The yellow arrow highlights the wave-vector of the incident light; the dashed lines highlight the collection angle determined by the numerical aperture of the objective lens used in experiments (see also the diagram in Fig.\ref{fgr:exp} (e) and (f)). (f) Forward to backward scattering intensity ratio as a function of wavelength.}
\label{fgr:polar}
\end{figure}

For a deeper understanding of the nature and appearance (e.g. the colour) of the different resonant modes found within a particle, we model the scattering spectra and corresponding angular distribution in the far field for $s$ and $p$ polarizations (i.e. with the electric field perpendicular and parallel to the scattering plain, respectively). We take into account the case of a TiO$_2$ sphere having r = 150 nm (Fig.\ref{fgr:polar} and fourth panel from the bottom in Fig.\ref{fgr:exp}(a)-(d)). Within the collection angle allowed by the numerical aperture of the microscope objective lens in use (highlighted by dashed lines in the polar plot in Fig.\ref{fgr:polar}(b)-(e)), the spectra of $s$ and $p$ polarizations are rather different (respectively blue and red spectra in Fig.\ref{fgr:polar}(a)). The relatively large angular aperture and intensity of the first two resonances ($\lambda$ $=$ 656 nm and $\lambda$ $=$ 486 nm, Fig.\ref{fgr:polar} (d) and (e)) lead to a better coupling within the NA of the microscope  with respect to higher order modes, and thus determine the main colour registered in our spectra and seen in dark field images. Higher order modes ($\lambda$ $=$ 377 nm and $\lambda$ $=$ 324 nm, Fig.\ref{fgr:polar} (b) and (c)) show a more complex behavior characterized by an overall larger angular aperture and the onset of multiple lobes specific of each polarization channel.

From the same analysis it is possible to assess the overall properties of the scattering intensity in the far field. Interestingly, for the first two resonances it is mostly forward-like for both $s$ and $p$ polarizations and it shows lobes with rather similar angular apertures, which corresponds to an almost isotropic behavior. Thus, we evaluate the forward to backward ratio (R = FW/BW) averaging the two polarization channels (Fig.\ref{fgr:polar} f)): in the investigated spectral range, R stays above 1 and it features strong fluctuations. In particular, a peak of R$\sim$9 is observed at about 740 nm.

\begin{figure}[t!]
\centering
\includegraphics[width=0.7\textwidth]{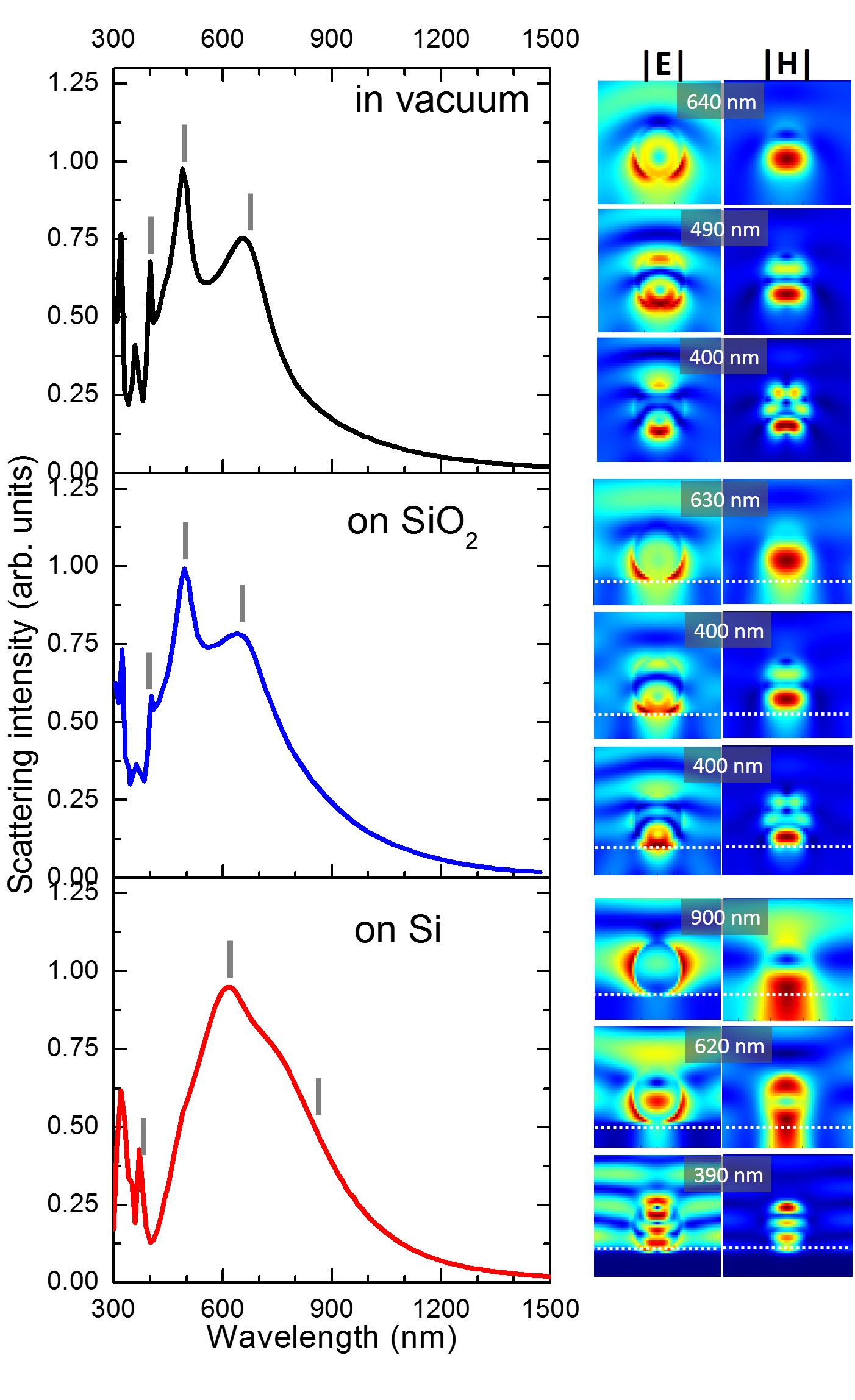}
\caption{FDTD simulations of the total far-field scattering intensity of a TiO$_{2}$ particle having a radius of 150 nm. Respectively from the top to the bottom panel are shown the cases of the same particle in vacuum, deposited on a silica substrate and on a silicon substrate. For each panel on the left, at the wavelengths highlighted by the gray tips, are shown on the right the intensity of the corresponding electric ($|E|$) and magnetic field ($|H|$).}
\label{fgr:FDTD}
\end{figure}

These theoretical predictions show a good directivity and polarization isotropy of the first two resonances (respectively ascribed to magnetic and electric dipolar modes\cite{Van2013,Markovich}, as also accounted for by FDTD simulations, Fig.\ref{fgr:FDTD}). The large FW/BW value found in theory at 740 nm for a Titania sphere, supports the idea of a Huygens-like source arising from the mutual interference of magnetic and electric dipolar modes (similar to the first Kerker scattering condition\cite{Kerker1983}) obtained with a dielectric and non-magnetic material\cite{Gomez2011,Krasnok2011,Person2013,Fu2013,Shibanuma2016}. This theoretical result is in good agreement with previous predictions for a similar system\cite{Zhang2015}: zero backward-scattering was predicted at about 800 nm for a TiO$_2$ particle having  r $=$ 150 nm and featuring a refractive index of about 2.5. The difference between our findings and those reported in \cite{Zhang2015} can be ascribed to the slightly lower refractive index used here.

These  features are important in view of the exploitation of these particles as low-loss, isotropic meta-materials with a magnetic response\cite{Popa2008,Evlyukhin2012} at frequencies not accessible by Si and Ge (e.g. for engineering their coupling with near-UV and blue LEDs based on GaN or in general with high-n substrates\cite{Van2013,Markovich}).
However, in spite of the reduced size of the contact point between a sphere and a plane, the presence of a high refractive index substrate underneath the particles may affect their resonances. Thus, in order to address this point, we performed Finite Difference Time Domain simulations (FDTD) of the total scattering power from a Titania sphere (r $=$ 150 nm) under plane wave illumination, at normal incidence with respect to the sample surface (the details of the FDTD method are provided in the devoted Section Theoretical models). Furthermore, we performed spectroscopic measurements of the light scattering in dark-field configuration (as those previously shown in Fig.\ref{fgr:exp}) on Titania spheres deposited on silica and on PDMS and compared them with those obtained on silicon (Fig.\ref{fgr:FDTD} and \ref{fgr:Si_vs_PDMS}). 

The comparison between the simulated far-field scattering spectra of a TiO$_{2}$ particle in vacuum and on silica shows a limited effect of the substrate on the resonances (Fig.\ref{fgr:FDTD} top and central panels): both cases show well defined resonances at similar wavelengths and having similar near filed intensity distribution of the electric and magnetic fields (respectively $|E|$ and $|H|$, right panels in Fig.\ref{fgr:FDTD}). From these near-field distributions we ascribe the resonance at $\sim$640 nm, $\sim$490 nm and $\sim$400 nm respectively to the magnetic dipolar mode, electric dipolar mode and magnetic quadrupolar mode, in agreement with simulations performed on a similar system\cite{Van2013,Markovich}. 

Comparing these FDTD simulations with the analytical theory (Fig.\ref{fgr:exp} d), fourth panel from the bottom) we observe a good agreement in the peak positions, whereas the different relative intensities and broadenings are ascribed to the different collection geometries used in FDTD simulations (total scattering in all the directions) and in the analytical model (scattering collected within the NA of the objective lens).

The same FDTD simulation run for a Titania sphere on a Si substrate shows a different picture (Fig.\ref{fgr:FDTD}, bottom panels): the two main peaks previously ascribed to magnetic electric and dipolar modes for the sphere on silica, now appear red-shifted, broadened and spectrally overlapped. Moreover, their near-field distributions present a relevant coupling within the substrate. These effects were predicted for spheres and cylinders made of silicon atop a high-refractive index substrate\cite{Spinelli,Van2013}. However, for TiO$_{2}$ spheres, we remark a more pronounced coupling of the magnetic dipolar mode in the underlying silicon substrate with respect to a Si sphere\cite{Van2013}. We ascribe this difference to the reduced refractive index contrast of a TiO$_{2}$ sphere on silicon with respect to one made of Si.

Higher order modes at short wavelengths ($\lambda$ $<$ 450 nm) show similar intensity  and broadening  for all the three FDTD simulations (Fig.\ref{fgr:FDTD}). Even if from the near-field features we cannot attribute the peak at 390 nm observed for the sphere on Si (Fig.\ref{fgr:FDTD} bottom panel) to the magnetic quadrupole peak found at  400 nm for the counterpart on silica (Fig.\ref{fgr:FDTD} central panel), we observe that in both cases the coupling with the substrate is quite limited. In fact, $|E|$ and $|H|$ stay well confined within the particle and the corresponding resonances are rather sharp.

Experimental spectra of light scattering of the same TiO$_{2}$ particle having a radius of about 120 nm, were first measured on a silicon substrate and then, after transfer, on a PDMS slice (Fig.\ref{fgr:Si_vs_PDMS} a)). Owing to the different refractive index of the underlying substrates, the spectrum changes as predicted by the FDTD simulations (Fig.\ref{fgr:FDTD}): from a broad peak at $\sim$550 nm on Si, two distinct peaks spring on PDMS, the first one at $\sim$450 nm and the second one at $\sim$500 nm.

Finally we monitor the behavior of higher order modes visible at short wavelength. To this end, we select two distinct, large particles featuring similar colours in dark-field images (r $>$ 200 nm, Fig.\ref{fgr:Si_vs_PDMS} b)). In both cases a sharp resonance is visible at about 450 nm, confirming the theoretical predictions obtained in FDTD simulations and the robustness of these Mie modes against a low refractive index contrast with the substrate: the mode confinement within the spheres can be large enough to prevent a complete quenching of the resonance which can exhibit a Q factor exceeding 20 as confirmed by a systematic investigation of large particles on Si and SiO $_{2}$ (Fig.\ref{fgr:Si_vs_PDMS}, central inset). 


In the perspective of using this aerosol approach for building structural colors, we show the changes in colour response evaluating the 1931 CIE chromaticity coefficients\cite{Smith1931} for the scattering in dark-field for smaller investigated particles (that are the cases of spheres whose spectrum lies in the visible range, r from 96 to 150 nm in Fig.\ref{fgr:exp}, as colorimetry functions are defined between 380 and 780 nm). In order to asses the scattering colours the spectra are not normalized to the illumination lamp.  In spite of the presence of the high refractive index Si substrate, our Mie resonators show neat colours in the visible range, spanning from blue-green up to orange of the CIE chromaticity gamut. Thus, even if TiO$_{2}$ has a lower refractive index with respect to Si and Ge, Titania spheres have the potential to be exploited for coloured metasurfaces\cite{Bottein,Sun2017} in analogy  with the group IV counterparts\cite{Devlin2016,Yue2017,Vashistha2017,Park2017,Zhu2017,Wood2017,Neder2017,Flauraud2017}.

\section{Conclusions}

In conclusion we showed that high throughput chemical methods, such as aerosol spray, can be used to form Titania-based, sub-micrometric spheres featuring low losses and relatively large refractive index up to blue and near-UV frequencies not accessible by conventional IV-IV compounds. The bright colours at visible frequencies are ascribed to Mie resonances formed within individual spheres, as confirmed by a systematic comparison of the experimental scattering spectra with the theoretical ones. 

Theoretical modeling highlights the possibility to obtain a negligible back-scattering, supporting the idea of a Huygens-like source based on individual spheres and an effective isotropic meta-material towards Yagi-Uda nano-antennas and efficient beam steering in the blue and near-UV frequencies, not accessible with conventional dissipative Si- and Ge-based systems. Nonetheless, the same aerosol method could be exploited for the fabrication of larger particles (>1 $\mu$m) sustaining whispering gallery modes, for light manipulation with photonic jets\cite{Chen2004} or efficient light transport in chains of dielectric elements\cite{Chen2006}. 

Finally, it is worth mentioning that the versatility of the aerosol technique in use may be adapted for a plethora of applications, as it allows for a direct transformation of the precursors solution in ready-to-use Mie resonators to be sprayed on a surface in a few seconds.  In this work, aerosol spray was well adapted to the simultaneous preparation of spheres with very different diameters. The poly-disperse size distribution of the particles would lead to an overall broad-band response of the ensemble, which may be adapted for the production of a structural white paint. However, when prepared with narrow diameter distribution they could be randomly assembled on a substrate by spraying them, to create a specific metasurface featuring structural colours. Several chemical-related techniques, such as electrospray\cite{Ding2005,Hong2008,Bock2011,Bock2012}, controlled sol-gel colloidal nucleation growth\cite{KES2002,LI2010} or microfluidic\cite{SHI2009} can be exploited to produce TiO$_{2}$ particles with much narrower size distribution. Besides, such particles suspension may be used as ink to draw coloured pictures by an ink-jet approach, as demonstrated with PMMA particles on fabrics\cite{LIU2017}. Since PMMA has a much lower refractive index than TiO$_{2}$, we believe that Titania spheres will be much more appropriated to produce structurally-colored images over large scale substrates.


\section{Experimental Section}

All chemicals were obtained from Aldrich and used as received. 
The spherical particles were prepared by atomizing a hydro-alcoholic solution containing 1:TiCl$_4$, 20:EtOH and 5:H$_2$O (molar fractions), in a carrying air flux using a TOPAZ ATM 210 aerosol generator (Fig.\ref{fgr:chem}).  

The nominal size of the droplets varies between 0.1 to several microns. The suspended micro-droplets/air mixture passed within a circular furnace at 300 $^{\circ}$C for a few seconds to complete evaporation and pre-stabilization of the dry particles (Fig.\ref{fgr:chem}). The suspended particles/air mixture was then accelerated at the furnace output using a nozzle and impacted over the target surface (e.g. a Si wafer or a microscope glass slide). This last deposition step lasted for about 10 sec preventing the clustering of the particles and keeping their density sufficiently low, thus allowing for spectroscopic investigation of single objects. The Si wafer sample was then heated at 600 $^{\circ}$C for 10 min for densification and crystallization of the spheres. In these configuration droplets of different diameters are generated, which is ideal for the present investigation. Titania particles were transfered from Si to PDMS by gently pressing the PDMS slide for a few seconds on the silicon surface and then pealed off. A fraction of the spheres were then randomly transfered on the low refractive index PDMS and recognized using predefined marks. Unfortunately the same process is not possible for the SiO$_{2}$ due to technical difficulties (PDMS is sticky whereas silica is not). Thus a direct comparison of the same spheres on Si and on SiO$_{2}$ was not possible

In parallel, the same solution was dip-coated using an ACEdip equipment from Solgelway to prepare a homogeneous thin layer on a Si wafer. The as-prepared coating was thermally treated in the same conditions, and analyzed by spectroscopic ellipsometry (Woollam M2000V) to extract the (n, k) dispersion used for the analytical model.

Grazing incidence X-ray Diffraction (GIXRD) investigation was performed in order to assess composition and crystallinity of the spheres. To this end, the spheres were deposited on a Si wafer for several minutes in order to increase their density without affecting their morphology and heated at 600 $^{\circ}$C for 10 min for densification and crystallization.  GIXRD patterns were recorded on a conventional diffractometer (PANalytical Empyrean) using Cu radiation ($\lambda$ = 0.154 nm), a rapid detector (PANalytical PIXcel) and a parallel plate collimator 0.27 degrees. The incident angle $\omega$ was 1.5 degrees (grazing incidence) and the in-plane angle measured from the Si substrate [010] direction was $\phi$ = 45 degrees.

The diffraction peaks were fitted by a Gauss profile function to measure the full width at half maximum (FWHM) of the peaks. In grazing incidence configuration the average crystallite size D (D is the size of coherent diffracting domains) can be extracted from the FWHM corrected (H$_c$) from the broadening due to the width of the beam footprint on the sample (H$_i$), and taking into account the experimental set-up (primary slit in the diffraction plane, goniometer radius and incident angle): H$_c$ = $\sqrt{H^2 - {H_i}^2}$, with H the measured FWHM \cite{Simeone2013}: D = 0.89$\lambda$/(H$_c \cos (\vartheta)$), where $\vartheta$  is the Bragg angle.

Observations by scanning electron microscopy (SEM) were performed on a FEI Helios 600 NanoLab. Micrographs were acquired using a  Through-the-Lens Detector (TLD) secondary electron detector, with a 5 kV acceleration voltage, a probe current of 0.17 nA and a working distance of 4.2 mm.

Spectroscopic measurements were performed on individual spheres to investigate their scattering properties. The spectra were collected by using an optical microscope (LEICA DMI 5000M) mounting a 100$\times$ magnification objective lens (numerical aperture NA = 0.75) in dark field configuration, coupled with a spectrometer and Si-based CCD linear array (Flame-T-VIS-NIR by Ocean Optics). In order to investigate individual spheres, the scattered light coming from each of them was collected using an optical fibre (Ocean Optics multimode fibre, VIS-NIR, core diameter 200 $\mu$m). 

Spectra from spheres of different radii were analyzed and compared with the expected results from analytical modeling. To this end, the raw experimental spectra were normalized to the excitation light by taking as a reference the spectrum of the light scattered from a dust spot on the sample surface. 

\section{Theoretical models}

\subsection{Analytical calculations}

Theoretical modeling of the spectral properties of the Titania particles were obtained from analytical solutions of Mie scattering from a sphere. They were obtained with an open source code\cite{Mie_code}. Excitation and collection geometries were adapted to those used in experiments (Fig.\ref{fgr:exp}(e) and (f)). Optical constants (n, k) and particles size used in theory were respectively deduced from ellipsometry and high-resolution SEM images.

\subsection{Finite Difference Time Domain simulations}

Finite Difference Time Domain (FDTD) were performed using a commercial software (Lumerical). The space surrounding the particles was discretized in a 3D grid (resolution 15 nm). The resonant particles were placed within a high accuracy region with twice the spatial resolution. The dimensions of the simulation space were 1$\times$1$\times$1 $\mu$m (in x, y and z respectively). The simulation space (substrate in the $<xy>$ plane) used periodic boundary conditions on the $x$ and $y$ extrema and absorbing boundary conditions in $z$. 

Light at normal incidence on $<xy>$ plane, was injected using a broadband plane wave source (source type total field-scattered field). The plane wave was polarized such that the electric field was oriented along the $x$ axis and the magnetic field along the $y$ axis. The near field distributions were collected by frequency domain monitors in the $<xz>$ plane (electric field) and $<yz>$ plane (magnetic field). 

The electromagnetic energy radiated from the resonators was collected by a
Poynting monitor box covering the al the space surrounding them. 

The refractive index used for the Titania composing the particles is that one obtained from the experimental ellipsometric measurements.

\section{Conflicts of interest}
There are no conflicts to declare.

\section{Acknowledgements}

This project has received funding from the European Unions Horizon 2020 research and innovation programme under grant agreement Laserlab-Europe ARES (no. 654148). The authors 
acknowledge the project PRCI network ULYSSES (ANR-15-CE24-0027-01) funded by the
French ANR agency; the SATT-Sud Est Project PROMETHEUS; the funding by A*MIDEX (reference. ANR- 11-IDEX-0001-02); the facilities of the NANOTECMAT platform at the IM2NP and of the microscopy center CP2M of Aix-Marseille University.


\bibliography{TiO2.bib} 

\providecommand*{\mcitethebibliography}{\thebibliography}
\csname @ifundefined\endcsname{endmcitethebibliography}
{\let\endmcitethebibliography\endthebibliography}{}
\begin{mcitethebibliography}{78}
\providecommand*{\natexlab}[1]{#1}
\providecommand*{\mciteSetBstSublistMode}[1]{}
\providecommand*{\mciteSetBstMaxWidthForm}[2]{}
\providecommand*{\mciteBstWouldAddEndPuncttrue}
  {\def\EndOfBibitem{\unskip.}}
\providecommand*{\mciteBstWouldAddEndPunctfalse}
  {\let\EndOfBibitem\relax}
\providecommand*{\mciteSetBstMidEndSepPunct}[3]{}
\providecommand*{\mciteSetBstSublistLabelBeginEnd}[3]{}
\providecommand*{\EndOfBibitem}{}
\mciteSetBstSublistMode{f}
\mciteSetBstMaxWidthForm{subitem}
{(\emph{\alph{mcitesubitemcount}})}
\mciteSetBstSublistLabelBeginEnd{\mcitemaxwidthsubitemform\space}
{\relax}{\relax}

\bibitem[Brongersma \emph{et~al.}(2014)Brongersma, Cui, and
  Fan]{Brongersma2014}
M.~L. Brongersma, Y.~Cui and S.~Fan, \emph{{N}at. {M}ater.}, 2014, \textbf{13},
  451--460\relax
\mciteBstWouldAddEndPuncttrue
\mciteSetBstMidEndSepPunct{\mcitedefaultmidpunct}
{\mcitedefaultendpunct}{\mcitedefaultseppunct}\relax
\EndOfBibitem
\bibitem[Kuznetsov \emph{et~al.}(2016)Kuznetsov, Miroshnichenko, Brongersma,
  Kivshar, and Luk’yanchuk]{Kuznetsov2016}
A.~Kuznetsov, A.~E. Miroshnichenko, M.~L. Brongersma, Y.~S. Kivshar and
  B.~Luk’yanchuk, \emph{{S}cience}, 2016, \textbf{354}, 846--853\relax
\mciteBstWouldAddEndPuncttrue
\mciteSetBstMidEndSepPunct{\mcitedefaultmidpunct}
{\mcitedefaultendpunct}{\mcitedefaultseppunct}\relax
\EndOfBibitem
\bibitem[S. and Z.(2016)]{jahani2016}
J.~S. and J.~Z., \emph{{N}at. {N}anotechnol.}, 2016, \textbf{11}, 23--36\relax
\mciteBstWouldAddEndPuncttrue
\mciteSetBstMidEndSepPunct{\mcitedefaultmidpunct}
{\mcitedefaultendpunct}{\mcitedefaultseppunct}\relax
\EndOfBibitem
\bibitem[Fenollosa \emph{et~al.}(2008)Fenollosa, Meseguer, and
  Tymczenko]{FEN2008}
R.~Fenollosa, F.~Meseguer and M.~Tymczenko, \emph{Advanced materials}, 2008,
  \textbf{20}, 95--98\relax
\mciteBstWouldAddEndPuncttrue
\mciteSetBstMidEndSepPunct{\mcitedefaultmidpunct}
{\mcitedefaultendpunct}{\mcitedefaultseppunct}\relax
\EndOfBibitem
\bibitem[Vynck \emph{et~al.}(2009)Vynck, Felbacq, Centeno, C{\u{a}}buz,
  Cassagne, and Guizal]{Vynck2009}
K.~Vynck, D.~Felbacq, E.~Centeno, A.~C{\u{a}}buz, D.~Cassagne and B.~Guizal,
  \emph{Physical review letters}, 2009, \textbf{102}, 133901\relax
\mciteBstWouldAddEndPuncttrue
\mciteSetBstMidEndSepPunct{\mcitedefaultmidpunct}
{\mcitedefaultendpunct}{\mcitedefaultseppunct}\relax
\EndOfBibitem
\bibitem[Evlyukhin \emph{et~al.}(2010)Evlyukhin, Reinhardt, Seidel,
  Luk’yanchuk, and Chichkov]{Evlyukhin2010}
A.~B. Evlyukhin, C.~Reinhardt, A.~Seidel, B.~S. Luk’yanchuk and B.~N.
  Chichkov, \emph{Physical Review B}, 2010, \textbf{82}, 045404\relax
\mciteBstWouldAddEndPuncttrue
\mciteSetBstMidEndSepPunct{\mcitedefaultmidpunct}
{\mcitedefaultendpunct}{\mcitedefaultseppunct}\relax
\EndOfBibitem
\bibitem[Garc{\'\i}a-Etxarri \emph{et~al.}(2011)Garc{\'\i}a-Etxarri,
  G{\'o}mez-Medina, Froufe-P{\'e}rez, L{\'o}pez, Chantada, Scheffold, Aizpurua,
  Nieto-Vesperinas, and S{\'a}enz]{Garcia2011}
A.~Garc{\'\i}a-Etxarri, R.~G{\'o}mez-Medina, L.~S. Froufe-P{\'e}rez,
  C.~L{\'o}pez, L.~Chantada, F.~Scheffold, J.~Aizpurua, M.~Nieto-Vesperinas and
  J.~J. S{\'a}enz, \emph{Optics express}, 2011, \textbf{19}, 4815--4826\relax
\mciteBstWouldAddEndPuncttrue
\mciteSetBstMidEndSepPunct{\mcitedefaultmidpunct}
{\mcitedefaultendpunct}{\mcitedefaultseppunct}\relax
\EndOfBibitem
\bibitem[Van~de Groep and Polman(2013)]{Van2013}
J.~Van~de Groep and A.~Polman, \emph{Optics express}, 2013, \textbf{21},
  26285--26302\relax
\mciteBstWouldAddEndPuncttrue
\mciteSetBstMidEndSepPunct{\mcitedefaultmidpunct}
{\mcitedefaultendpunct}{\mcitedefaultseppunct}\relax
\EndOfBibitem
\bibitem[Staude and Schilling(2017)]{Staude2017}
I.~Staude and J.~Schilling, \emph{Nature Photonics}, 2017, \textbf{11},
  274\relax
\mciteBstWouldAddEndPuncttrue
\mciteSetBstMidEndSepPunct{\mcitedefaultmidpunct}
{\mcitedefaultendpunct}{\mcitedefaultseppunct}\relax
\EndOfBibitem
\bibitem[She \emph{et~al.}(2018)She, Zhang, Shian, Clarke, and
  Capasso]{SHE2018}
A.~She, S.~Zhang, S.~Shian, D.~R. Clarke and F.~Capasso, \emph{Science
  Advances}, 2018, \textbf{4}, eaap9957\relax
\mciteBstWouldAddEndPuncttrue
\mciteSetBstMidEndSepPunct{\mcitedefaultmidpunct}
{\mcitedefaultendpunct}{\mcitedefaultseppunct}\relax
\EndOfBibitem
\bibitem[Gomez-Medina \emph{et~al.}(2011)Gomez-Medina, Garcia-Camara,
  Su{\'a}rez-Lacalle, Gonz{\'a}lez, Moreno, Nieto-Vesperinas, and
  S{\'a}enz]{Gomez2011}
R.~Gomez-Medina, B.~Garcia-Camara, I.~Su{\'a}rez-Lacalle, F.~Gonz{\'a}lez,
  F.~Moreno, M.~Nieto-Vesperinas and J.~J. S{\'a}enz, \emph{Journal of
  Nanophotonics}, 2011, \textbf{5}, 053512\relax
\mciteBstWouldAddEndPuncttrue
\mciteSetBstMidEndSepPunct{\mcitedefaultmidpunct}
{\mcitedefaultendpunct}{\mcitedefaultseppunct}\relax
\EndOfBibitem
\bibitem[Zhu \emph{et~al.}(2017)Zhu, Yan, Levy, Mortensen, and
  Kristensen]{Zhu2017}
X.~Zhu, W.~Yan, U.~Levy, N.~A. Mortensen and A.~Kristensen, \emph{Science
  Advances}, 2017, \textbf{3}, e1602487\relax
\mciteBstWouldAddEndPuncttrue
\mciteSetBstMidEndSepPunct{\mcitedefaultmidpunct}
{\mcitedefaultendpunct}{\mcitedefaultseppunct}\relax
\EndOfBibitem
\bibitem[Wood \emph{et~al.}(2017)Wood, Naffouti, Berthelot, David, Claude,
  M{\'e}tayer, Delobbe, Favre, Ronda, Berbezier,\emph{et~al.}]{Wood2017}
T.~Wood, M.~Naffouti, J.~Berthelot, T.~David, J.-B. Claude, L.~M{\'e}tayer,
  A.~Delobbe, L.~Favre, A.~Ronda, I.~Berbezier \emph{et~al.}, \emph{ACS
  photonics}, 2017, \textbf{4}, 873--883\relax
\mciteBstWouldAddEndPuncttrue
\mciteSetBstMidEndSepPunct{\mcitedefaultmidpunct}
{\mcitedefaultendpunct}{\mcitedefaultseppunct}\relax
\EndOfBibitem
\bibitem[Bouabdellaoui \emph{et~al.}(2018)Bouabdellaoui, Checcucci, Wood,
  Naffouti, Sena, Liu, Ruiz, Duche, le~Rouzo, Escoubas,\emph{et~al.}]{BOU2018}
M.~Bouabdellaoui, S.~Checcucci, T.~Wood, M.~Naffouti, R.~P. Sena, K.~Liu, C.~M.
  Ruiz, D.~Duche, J.~le~Rouzo, L.~Escoubas \emph{et~al.}, \emph{Physical Review
  Materials}, 2018, \textbf{2}, 035203\relax
\mciteBstWouldAddEndPuncttrue
\mciteSetBstMidEndSepPunct{\mcitedefaultmidpunct}
{\mcitedefaultendpunct}{\mcitedefaultseppunct}\relax
\EndOfBibitem
\bibitem[Philipp and Taft(1959)]{PHI1959}
H.~Philipp and E.~Taft, \emph{Physical Review}, 1959, \textbf{113}, 1002\relax
\mciteBstWouldAddEndPuncttrue
\mciteSetBstMidEndSepPunct{\mcitedefaultmidpunct}
{\mcitedefaultendpunct}{\mcitedefaultseppunct}\relax
\EndOfBibitem
\bibitem[Philipp and Taft(1960)]{PHI1960}
H.~Philipp and E.~Taft, \emph{Physical Review}, 1960, \textbf{120}, 37\relax
\mciteBstWouldAddEndPuncttrue
\mciteSetBstMidEndSepPunct{\mcitedefaultmidpunct}
{\mcitedefaultendpunct}{\mcitedefaultseppunct}\relax
\EndOfBibitem
\bibitem[Gar{\'\i}n \emph{et~al.}(2014)Gar{\'\i}n, Fenollosa, Alcubilla, Shi,
  Marsal, and Meseguer]{Garin2014}
M.~Gar{\'\i}n, R.~Fenollosa, R.~Alcubilla, L.~Shi, L.~Marsal and F.~Meseguer,
  \emph{Nature communications}, 2014, \textbf{5}, 3440\relax
\mciteBstWouldAddEndPuncttrue
\mciteSetBstMidEndSepPunct{\mcitedefaultmidpunct}
{\mcitedefaultendpunct}{\mcitedefaultseppunct}\relax
\EndOfBibitem
\bibitem[Shi \emph{et~al.}(2012)Shi, Tuzer, Fenollosa, and Meseguer]{Shi2012}
L.~Shi, T.~U. Tuzer, R.~Fenollosa and F.~Meseguer, \emph{Advanced materials},
  2012, \textbf{24}, 5934--5938\relax
\mciteBstWouldAddEndPuncttrue
\mciteSetBstMidEndSepPunct{\mcitedefaultmidpunct}
{\mcitedefaultendpunct}{\mcitedefaultseppunct}\relax
\EndOfBibitem
\bibitem[Abbarchi \emph{et~al.}(2014)Abbarchi, Naffouti, Vial, Benkouider,
  Lermusiaux, Favre, Ronda, Bidault, Berbezier, and Bonod]{Abbarchi2014}
M.~Abbarchi, M.~Naffouti, B.~Vial, A.~Benkouider, L.~Lermusiaux, L.~Favre,
  A.~Ronda, S.~Bidault, I.~Berbezier and N.~Bonod, \emph{ACS nano}, 2014,
  \textbf{8}, 11181--11190\relax
\mciteBstWouldAddEndPuncttrue
\mciteSetBstMidEndSepPunct{\mcitedefaultmidpunct}
{\mcitedefaultendpunct}{\mcitedefaultseppunct}\relax
\EndOfBibitem
\bibitem[Naffouti \emph{et~al.}(2016)Naffouti, David, Benkouider, Favre, Cabie,
  Ronda, Berbezier, and Abbarchi]{Naffouti2016}
M.~Naffouti, T.~David, A.~Benkouider, L.~Favre, M.~Cabie, A.~Ronda,
  I.~Berbezier and M.~Abbarchi, \emph{Nanotechnology}, 2016, \textbf{27},
  305602\relax
\mciteBstWouldAddEndPuncttrue
\mciteSetBstMidEndSepPunct{\mcitedefaultmidpunct}
{\mcitedefaultendpunct}{\mcitedefaultseppunct}\relax
\EndOfBibitem
\bibitem[Naffouti \emph{et~al.}(2016)Naffouti, David, Benkouider, Favre, Ronda,
  Berbezier, Bidault, Bonod, and Abbarchi]{Naffouti2016c}
M.~Naffouti, T.~David, A.~Benkouider, L.~Favre, A.~Ronda, I.~Berbezier,
  S.~Bidault, N.~Bonod and M.~Abbarchi, \emph{Nanoscale}, 2016, \textbf{8},
  2844--2849\relax
\mciteBstWouldAddEndPuncttrue
\mciteSetBstMidEndSepPunct{\mcitedefaultmidpunct}
{\mcitedefaultendpunct}{\mcitedefaultseppunct}\relax
\EndOfBibitem
\bibitem[Naffouti \emph{et~al.}(2016)Naffouti, David, Benkouider, Favre,
  Delobbe, Ronda, Berbezier, and Abbarchi]{Naffouti2016b}
M.~Naffouti, T.~David, A.~Benkouider, L.~Favre, A.~Delobbe, A.~Ronda,
  I.~Berbezier and M.~Abbarchi, \emph{Small}, 2016, \textbf{12},
  6115--6123\relax
\mciteBstWouldAddEndPuncttrue
\mciteSetBstMidEndSepPunct{\mcitedefaultmidpunct}
{\mcitedefaultendpunct}{\mcitedefaultseppunct}\relax
\EndOfBibitem
\bibitem[Naffouti \emph{et~al.}(2017)Naffouti, Backofen, Salvalaglio, Bottein,
  Lodari, Voigt, David, Benkouider, Fraj, Favre,\emph{et~al.}]{Naffouti2017}
M.~Naffouti, R.~Backofen, M.~Salvalaglio, T.~Bottein, M.~Lodari, A.~Voigt,
  T.~David, A.~Benkouider, I.~Fraj, L.~Favre \emph{et~al.}, \emph{Science
  advances}, 2017, \textbf{3}, eaao1472\relax
\mciteBstWouldAddEndPuncttrue
\mciteSetBstMidEndSepPunct{\mcitedefaultmidpunct}
{\mcitedefaultendpunct}{\mcitedefaultseppunct}\relax
\EndOfBibitem
\bibitem[Kim(1996)]{Kim1996}
S.~Kim, \emph{Applied optics}, 1996, \textbf{35}, 6703--6707\relax
\mciteBstWouldAddEndPuncttrue
\mciteSetBstMidEndSepPunct{\mcitedefaultmidpunct}
{\mcitedefaultendpunct}{\mcitedefaultseppunct}\relax
\EndOfBibitem
\bibitem[Siefke \emph{et~al.}(2016)Siefke, Kroker, Pfeiffer, Puffky, Dietrich,
  Franta, Ohl{\'\i}dal, Szeghalmi, Kley, and T{\"u}nnermann]{Siefke2016}
T.~Siefke, S.~Kroker, K.~Pfeiffer, O.~Puffky, K.~Dietrich, D.~Franta,
  I.~Ohl{\'\i}dal, A.~Szeghalmi, E.-B. Kley and A.~T{\"u}nnermann,
  \emph{Advanced Optical Materials}, 2016, \textbf{4}, 1780--1786\relax
\mciteBstWouldAddEndPuncttrue
\mciteSetBstMidEndSepPunct{\mcitedefaultmidpunct}
{\mcitedefaultendpunct}{\mcitedefaultseppunct}\relax
\EndOfBibitem
\bibitem[Khorasaninejad \emph{et~al.}(2016)Khorasaninejad, Chen, Devlin, Oh,
  Zhu, and Capasso]{Khorasaninejad2016}
M.~Khorasaninejad, W.~T. Chen, R.~C. Devlin, J.~Oh, A.~Y. Zhu and F.~Capasso,
  \emph{Science}, 2016, \textbf{352}, 1190--1194\relax
\mciteBstWouldAddEndPuncttrue
\mciteSetBstMidEndSepPunct{\mcitedefaultmidpunct}
{\mcitedefaultendpunct}{\mcitedefaultseppunct}\relax
\EndOfBibitem
\bibitem[Abarca \emph{et~al.}(2016)Abarca, G\'omez-Sal, Mart\'in, Mena, Poblet,
  and Y\'elamos]{Gutruf}
A.~Abarca, P.~G\'omez-Sal, A.~Mart\'in, M.~Mena, J.~M. Poblet and C.~Y\'elamos,
  \emph{{ACS} {N}ano}, 2016, \textbf{10}, 133--141\relax
\mciteBstWouldAddEndPuncttrue
\mciteSetBstMidEndSepPunct{\mcitedefaultmidpunct}
{\mcitedefaultendpunct}{\mcitedefaultseppunct}\relax
\EndOfBibitem
\bibitem[Devlin \emph{et~al.}(2016)Devlin, Khorasaninejad, Chen, Oh, and
  Capasso]{Devlin2016}
R.~C. Devlin, M.~Khorasaninejad, W.~T. Chen, J.~Oh and F.~Capasso,
  \emph{Proceedings of the National Academy of Sciences}, 2016, \textbf{113},
  10473--10478\relax
\mciteBstWouldAddEndPuncttrue
\mciteSetBstMidEndSepPunct{\mcitedefaultmidpunct}
{\mcitedefaultendpunct}{\mcitedefaultseppunct}\relax
\EndOfBibitem
\bibitem[Sun \emph{et~al.}(2017)Sun, Zhou, Zhang, Gao, Duan, Xiao, and
  Song]{Sun2017}
S.~Sun, Z.~Zhou, C.~Zhang, Y.~Gao, Z.~Duan, S.~Xiao and Q.~Song, \emph{ACS
  nano}, 2017, \textbf{11}, 4445--4452\relax
\mciteBstWouldAddEndPuncttrue
\mciteSetBstMidEndSepPunct{\mcitedefaultmidpunct}
{\mcitedefaultendpunct}{\mcitedefaultseppunct}\relax
\EndOfBibitem
\bibitem[Sun \emph{et~al.}(2018)Sun, Yang, Zhang, Jing, Gao, Yu, Song, and
  Xiao]{SUN2018}
S.~Sun, W.~Yang, C.~Zhang, J.~Jing, Y.~Gao, X.~Yu, Q.~Song and S.~Xiao,
  \emph{ACS nano}, 2018\relax
\mciteBstWouldAddEndPuncttrue
\mciteSetBstMidEndSepPunct{\mcitedefaultmidpunct}
{\mcitedefaultendpunct}{\mcitedefaultseppunct}\relax
\EndOfBibitem
\bibitem[Bottein \emph{et~al.}(2016)Bottein, Wood, David, Claude, Favre,
  Berbezier, Ronda, Abbarchi, and Grosso]{Bottein}
T.~Bottein, T.~Wood, T.~David, J.~Claude, L.~Favre, I.~Berbezier, A.~Ronda,
  M.~Abbarchi and D.~Grosso, \emph{Adv. Funct. Mater.}, 2016\relax
\mciteBstWouldAddEndPuncttrue
\mciteSetBstMidEndSepPunct{\mcitedefaultmidpunct}
{\mcitedefaultendpunct}{\mcitedefaultseppunct}\relax
\EndOfBibitem
\bibitem[Bottein \emph{et~al.}(2018)Bottein, Dalstein, Putero, Cattoni,
  Faustini, Abbarchi, and Grosso]{Bottein2018}
T.~Bottein, O.~Dalstein, M.~Putero, A.~Cattoni, M.~Faustini, M.~Abbarchi and
  D.~Grosso, \emph{Nanoscale}, 2018\relax
\mciteBstWouldAddEndPuncttrue
\mciteSetBstMidEndSepPunct{\mcitedefaultmidpunct}
{\mcitedefaultendpunct}{\mcitedefaultseppunct}\relax
\EndOfBibitem
\bibitem[Tzarouchis and Sihvola(2018)]{TZA2018}
D.~Tzarouchis and A.~Sihvola, \emph{Applied Sciences}, 2018, \textbf{8},
  184\relax
\mciteBstWouldAddEndPuncttrue
\mciteSetBstMidEndSepPunct{\mcitedefaultmidpunct}
{\mcitedefaultendpunct}{\mcitedefaultseppunct}\relax
\EndOfBibitem
\bibitem[Krasnok \emph{et~al.}(2011)Krasnok, Miroshnichenko, Belov, and
  Kivshar]{Krasnok2011}
A.~E. Krasnok, A.~E. Miroshnichenko, P.~A. Belov and Y.~S. Kivshar, \emph{JETP
  letters}, 2011, \textbf{94}, 593--598\relax
\mciteBstWouldAddEndPuncttrue
\mciteSetBstMidEndSepPunct{\mcitedefaultmidpunct}
{\mcitedefaultendpunct}{\mcitedefaultseppunct}\relax
\EndOfBibitem
\bibitem[Fu \emph{et~al.}(2013)Fu, Kuznetsov, Miroshnichenko, Yu, and
  Luk’yanchuk]{Fu2013}
Y.~H. Fu, A.~I. Kuznetsov, A.~E. Miroshnichenko, Y.~F. Yu and B.~Luk’yanchuk,
  \emph{Nature communications}, 2013, \textbf{4}, 1527\relax
\mciteBstWouldAddEndPuncttrue
\mciteSetBstMidEndSepPunct{\mcitedefaultmidpunct}
{\mcitedefaultendpunct}{\mcitedefaultseppunct}\relax
\EndOfBibitem
\bibitem[Campione \emph{et~al.}(2015)Campione, Basilio, Warne, and
  Sinclair]{CAM2015}
S.~Campione, L.~I. Basilio, L.~K. Warne and M.~B. Sinclair, \emph{Optics
  Express}, 2015, \textbf{23}, 2293--2307\relax
\mciteBstWouldAddEndPuncttrue
\mciteSetBstMidEndSepPunct{\mcitedefaultmidpunct}
{\mcitedefaultendpunct}{\mcitedefaultseppunct}\relax
\EndOfBibitem
\bibitem[Zhang \emph{et~al.}(2015)Zhang, Nieto-Vesperinas, and
  S{\'a}enz]{Zhang2015}
Y.~Zhang, M.~Nieto-Vesperinas and J.~J. S{\'a}enz, \emph{Journal of Optics},
  2015, \textbf{17}, 105612\relax
\mciteBstWouldAddEndPuncttrue
\mciteSetBstMidEndSepPunct{\mcitedefaultmidpunct}
{\mcitedefaultendpunct}{\mcitedefaultseppunct}\relax
\EndOfBibitem
\bibitem[Yan \emph{et~al.}(2015)Yan, Liu, Lin, Wang, Chen, Wang, and
  Yang]{Yan2015}
J.~Yan, P.~Liu, Z.~Lin, H.~Wang, H.~Chen, C.~Wang and G.~Yang, \emph{Nature
  communications}, 2015, \textbf{6}, 7042\relax
\mciteBstWouldAddEndPuncttrue
\mciteSetBstMidEndSepPunct{\mcitedefaultmidpunct}
{\mcitedefaultendpunct}{\mcitedefaultseppunct}\relax
\EndOfBibitem
\bibitem[Liberal \emph{et~al.}(2015)Liberal, Ederra, Gonzalo, and
  Ziolkowski]{Liberal2015}
I.~Liberal, I.~Ederra, R.~Gonzalo and R.~W. Ziolkowski, \emph{Journal of
  Optics}, 2015, \textbf{17}, 072001\relax
\mciteBstWouldAddEndPuncttrue
\mciteSetBstMidEndSepPunct{\mcitedefaultmidpunct}
{\mcitedefaultendpunct}{\mcitedefaultseppunct}\relax
\EndOfBibitem
\bibitem[Albella \emph{et~al.}(2015)Albella, Shibanuma, and Maier]{Albella2015}
P.~Albella, T.~Shibanuma and S.~A. Maier, \emph{Scientific reports}, 2015,
  \textbf{5}, 18322\relax
\mciteBstWouldAddEndPuncttrue
\mciteSetBstMidEndSepPunct{\mcitedefaultmidpunct}
{\mcitedefaultendpunct}{\mcitedefaultseppunct}\relax
\EndOfBibitem
\bibitem[Tribelsky \emph{et~al.}(2015)Tribelsky, Geffrin, Litman, Eyraud, and
  Moreno]{Tribelsky2015}
M.~I. Tribelsky, J.-M. Geffrin, A.~Litman, C.~Eyraud and F.~Moreno,
  \emph{Scientific reports}, 2015, \textbf{5}, 12288\relax
\mciteBstWouldAddEndPuncttrue
\mciteSetBstMidEndSepPunct{\mcitedefaultmidpunct}
{\mcitedefaultendpunct}{\mcitedefaultseppunct}\relax
\EndOfBibitem
\bibitem[Wo{\'z}niak \emph{et~al.}(2015)Wo{\'z}niak, Banzer, and
  Leuchs]{Wozniak2015}
P.~Wo{\'z}niak, P.~Banzer and G.~Leuchs, \emph{Laser \& Photonics Reviews},
  2015, \textbf{9}, 231--240\relax
\mciteBstWouldAddEndPuncttrue
\mciteSetBstMidEndSepPunct{\mcitedefaultmidpunct}
{\mcitedefaultendpunct}{\mcitedefaultseppunct}\relax
\EndOfBibitem
\bibitem[Shibanuma \emph{et~al.}(2016)Shibanuma, Albella, and
  Maier]{Shibanuma2016}
T.~Shibanuma, P.~Albella and S.~A. Maier, \emph{Nanoscale}, 2016, \textbf{8},
  14184--14192\relax
\mciteBstWouldAddEndPuncttrue
\mciteSetBstMidEndSepPunct{\mcitedefaultmidpunct}
{\mcitedefaultendpunct}{\mcitedefaultseppunct}\relax
\EndOfBibitem
\bibitem[Ollanik \emph{et~al.}(2018)Ollanik, Smith, Belue, and
  Escarra]{OLL2018}
A.~J. Ollanik, J.~A. Smith, M.~J. Belue and M.~D. Escarra, \emph{ACS
  Photonics}, 2018\relax
\mciteBstWouldAddEndPuncttrue
\mciteSetBstMidEndSepPunct{\mcitedefaultmidpunct}
{\mcitedefaultendpunct}{\mcitedefaultseppunct}\relax
\EndOfBibitem
\bibitem[Komar \emph{et~al.}(2018)Komar, Paniagua-Dom{\'\i}nguez,
  Miroshnichenko, Yu, Kivshar, Kuznetsov, and Neshev]{KOM2018}
A.~Komar, R.~Paniagua-Dom{\'\i}nguez, A.~Miroshnichenko, Y.~F. Yu, Y.~S.
  Kivshar, A.~I. Kuznetsov and D.~Neshev, \emph{ACS Photonics}, 2018\relax
\mciteBstWouldAddEndPuncttrue
\mciteSetBstMidEndSepPunct{\mcitedefaultmidpunct}
{\mcitedefaultendpunct}{\mcitedefaultseppunct}\relax
\EndOfBibitem
\bibitem[Krasnok \emph{et~al.}(2015)Krasnok, Maloshtan, Chigrin, Kivshar, and
  Belov]{Krasnok2015}
A.~E. Krasnok, A.~Maloshtan, D.~N. Chigrin, Y.~S. Kivshar and P.~A. Belov,
  \emph{Laser \& Photonics Reviews}, 2015, \textbf{9}, 385--391\relax
\mciteBstWouldAddEndPuncttrue
\mciteSetBstMidEndSepPunct{\mcitedefaultmidpunct}
{\mcitedefaultendpunct}{\mcitedefaultseppunct}\relax
\EndOfBibitem
\bibitem[Vaskin \emph{et~al.}(2018)Vaskin, Bohn, Chong, Bucher, Zilk, Choi,
  Neshev, Kivshar, Pertsch, and Staude]{VAS2018}
A.~Vaskin, J.~Bohn, K.~E. Chong, T.~Bucher, M.~Zilk, D.-Y. Choi, D.~N. Neshev,
  Y.~S. Kivshar, T.~Pertsch and I.~Staude, \emph{ACS Photonics}, 2018\relax
\mciteBstWouldAddEndPuncttrue
\mciteSetBstMidEndSepPunct{\mcitedefaultmidpunct}
{\mcitedefaultendpunct}{\mcitedefaultseppunct}\relax
\EndOfBibitem
\bibitem[Savelev \emph{et~al.}(2015)Savelev, Filonov, Petrov, Krasnok, Belov,
  and Kivshar]{Savelev2015}
R.~S. Savelev, D.~S. Filonov, M.~I. Petrov, A.~E. Krasnok, P.~A. Belov and
  Y.~S. Kivshar, \emph{Physical Review B}, 2015, \textbf{92}, 155415\relax
\mciteBstWouldAddEndPuncttrue
\mciteSetBstMidEndSepPunct{\mcitedefaultmidpunct}
{\mcitedefaultendpunct}{\mcitedefaultseppunct}\relax
\EndOfBibitem
\bibitem[Andres-Arroyo \emph{et~al.}(2016)Andres-Arroyo, Gupta, Wang, Gooding,
  and Reece]{Andres2016}
A.~Andres-Arroyo, B.~Gupta, F.~Wang, J.~J. Gooding and P.~J. Reece, \emph{Nano
  letters}, 2016, \textbf{16}, 1903--1910\relax
\mciteBstWouldAddEndPuncttrue
\mciteSetBstMidEndSepPunct{\mcitedefaultmidpunct}
{\mcitedefaultendpunct}{\mcitedefaultseppunct}\relax
\EndOfBibitem
\bibitem[Yan \emph{et~al.}(2015)Yan, Liu, Lin, Wang, Chen, Wang, and
  Yang]{Yan2015b}
J.~Yan, P.~Liu, Z.~Lin, H.~Wang, H.~Chen, C.~Wang and G.~Yang, \emph{Acs Nano},
  2015, \textbf{9}, 2968--2980\relax
\mciteBstWouldAddEndPuncttrue
\mciteSetBstMidEndSepPunct{\mcitedefaultmidpunct}
{\mcitedefaultendpunct}{\mcitedefaultseppunct}\relax
\EndOfBibitem
\bibitem[Tribelsky \emph{et~al.}(2016)Tribelsky, Geffrin, Litman, Eyraud, and
  Moreno]{Tribelsky2016}
M.~I. Tribelsky, J.-M. Geffrin, A.~Litman, C.~Eyraud and F.~Moreno,
  \emph{Physical Review B}, 2016, \textbf{94}, 121110\relax
\mciteBstWouldAddEndPuncttrue
\mciteSetBstMidEndSepPunct{\mcitedefaultmidpunct}
{\mcitedefaultendpunct}{\mcitedefaultseppunct}\relax
\EndOfBibitem
\bibitem[Coenen \emph{et~al.}(2013)Coenen, Groep, and Polman]{Coenen}
T.~Coenen, J.~V.~D. Groep and A.~Polman, \emph{ACS nano}, 2013, \textbf{7},
  1689--1698\relax
\mciteBstWouldAddEndPuncttrue
\mciteSetBstMidEndSepPunct{\mcitedefaultmidpunct}
{\mcitedefaultendpunct}{\mcitedefaultseppunct}\relax
\EndOfBibitem
\bibitem[N{\v{e}}mec \emph{et~al.}(2012)N{\v{e}}mec, Kadlec, Kadlec,
  Ku{\v{z}}el, Yahiaoui, Chung, Elissalde, Maglione, and Mounaix]{Nvemec2012}
H.~N{\v{e}}mec, C.~Kadlec, F.~Kadlec, P.~Ku{\v{z}}el, R.~Yahiaoui, U.-C. Chung,
  C.~Elissalde, M.~Maglione and P.~Mounaix, \emph{Applied Physics Letters},
  2012, \textbf{100}, 061117\relax
\mciteBstWouldAddEndPuncttrue
\mciteSetBstMidEndSepPunct{\mcitedefaultmidpunct}
{\mcitedefaultendpunct}{\mcitedefaultseppunct}\relax
\EndOfBibitem
\bibitem[Navarro-C{\'\i}a \emph{et~al.}(2013)Navarro-C{\'\i}a, Natrella,
  Dominec, Delagnes, Ku{\v{z}}el, Mounaix, Graham, Renaud, Seeds, and
  Mitrofanov]{Navarro2013}
M.~Navarro-C{\'\i}a, M.~Natrella, F.~Dominec, J.-C. Delagnes, P.~Ku{\v{z}}el,
  P.~Mounaix, C.~Graham, C.~Renaud, A.~Seeds and O.~Mitrofanov, \emph{Applied
  Physics Letters}, 2013, \textbf{103}, 221103\relax
\mciteBstWouldAddEndPuncttrue
\mciteSetBstMidEndSepPunct{\mcitedefaultmidpunct}
{\mcitedefaultendpunct}{\mcitedefaultseppunct}\relax
\EndOfBibitem
\bibitem[Spinelli \emph{et~al.}(2012)Spinelli, Verschuuren, and
  Polman]{Spinelli}
P.~Spinelli, M.~Verschuuren and A.~Polman, \emph{Nature Communications}, 2012,
  \textbf{3}, 692\relax
\mciteBstWouldAddEndPuncttrue
\mciteSetBstMidEndSepPunct{\mcitedefaultmidpunct}
{\mcitedefaultendpunct}{\mcitedefaultseppunct}\relax
\EndOfBibitem
\bibitem[Markovich \emph{et~al.}(2014)Markovich, Ginzburg, Samusev, Belov, and
  Zayats]{Markovich}
D.~L. Markovich, P.~Ginzburg, A.~K. Samusev, P.~A. Belov and A.~V. Zayats,
  \emph{Optics express}, 2014, \textbf{22}, 10693\relax
\mciteBstWouldAddEndPuncttrue
\mciteSetBstMidEndSepPunct{\mcitedefaultmidpunct}
{\mcitedefaultendpunct}{\mcitedefaultseppunct}\relax
\EndOfBibitem
\bibitem[Popa and Cummer(2008)]{Popa2008}
B.-I. Popa and S.~A. Cummer, \emph{Physical review letters}, 2008,
  \textbf{100}, 207401\relax
\mciteBstWouldAddEndPuncttrue
\mciteSetBstMidEndSepPunct{\mcitedefaultmidpunct}
{\mcitedefaultendpunct}{\mcitedefaultseppunct}\relax
\EndOfBibitem
\bibitem[Evlyukhin \emph{et~al.}(2012)Evlyukhin, Novikov, Zywietz, Eriksen,
  Reinhardt, Bozhevolnyi, and Chichkov]{Evlyukhin2012}
A.~B. Evlyukhin, S.~M. Novikov, U.~Zywietz, R.~L. Eriksen, C.~Reinhardt, S.~I.
  Bozhevolnyi and B.~N. Chichkov, \emph{Nano letters}, 2012, \textbf{12},
  3749--3755\relax
\mciteBstWouldAddEndPuncttrue
\mciteSetBstMidEndSepPunct{\mcitedefaultmidpunct}
{\mcitedefaultendpunct}{\mcitedefaultseppunct}\relax
\EndOfBibitem
\bibitem[Kerker \emph{et~al.}(1983)Kerker, Wang, and Giles]{Kerker1983}
M.~Kerker, D.-S. Wang and C.~Giles, \emph{JOSA}, 1983, \textbf{73},
  765--767\relax
\mciteBstWouldAddEndPuncttrue
\mciteSetBstMidEndSepPunct{\mcitedefaultmidpunct}
{\mcitedefaultendpunct}{\mcitedefaultseppunct}\relax
\EndOfBibitem
\bibitem[Person \emph{et~al.}(2013)Person, Jain, Lapin, Saenz, Wicks, and
  Novotny]{Person2013}
S.~Person, M.~Jain, Z.~Lapin, J.~J. Saenz, G.~Wicks and L.~Novotny, \emph{Nano
  letters}, 2013, \textbf{13}, 1806--1809\relax
\mciteBstWouldAddEndPuncttrue
\mciteSetBstMidEndSepPunct{\mcitedefaultmidpunct}
{\mcitedefaultendpunct}{\mcitedefaultseppunct}\relax
\EndOfBibitem
\bibitem[Smith and Guild(1931)]{Smith1931}
T.~Smith and J.~Guild, \emph{Transactions of the optical society}, 1931,
  \textbf{33}, 73\relax
\mciteBstWouldAddEndPuncttrue
\mciteSetBstMidEndSepPunct{\mcitedefaultmidpunct}
{\mcitedefaultendpunct}{\mcitedefaultseppunct}\relax
\EndOfBibitem
\bibitem[Yue \emph{et~al.}(2017)Yue, Gao, Lee, Kim, and Choi]{Yue2017}
W.~Yue, S.~Gao, S.-S. Lee, E.-S. Kim and D.-Y. Choi, \emph{Laser \& Photonics
  Reviews}, 2017, \textbf{11}, year\relax
\mciteBstWouldAddEndPuncttrue
\mciteSetBstMidEndSepPunct{\mcitedefaultmidpunct}
{\mcitedefaultendpunct}{\mcitedefaultseppunct}\relax
\EndOfBibitem
\bibitem[Vashistha \emph{et~al.}(2017)Vashistha, Vaidya, Gruszecki,
  Serebryannikov, and Krawczyk]{Vashistha2017}
V.~Vashistha, G.~Vaidya, P.~Gruszecki, A.~E. Serebryannikov and M.~Krawczyk,
  \emph{Scientific Reports}, 2017, \textbf{7}, 8092\relax
\mciteBstWouldAddEndPuncttrue
\mciteSetBstMidEndSepPunct{\mcitedefaultmidpunct}
{\mcitedefaultendpunct}{\mcitedefaultseppunct}\relax
\EndOfBibitem
\bibitem[Park \emph{et~al.}(2017)Park, Shrestha, Yue, Gao, Lee, Kim, and
  Choi]{Park2017}
C.-S. Park, V.~R. Shrestha, W.~Yue, S.~Gao, S.-S. Lee, E.-S. Kim and D.-Y.
  Choi, \emph{Scientific Reports}, 2017, \textbf{7}, 2556\relax
\mciteBstWouldAddEndPuncttrue
\mciteSetBstMidEndSepPunct{\mcitedefaultmidpunct}
{\mcitedefaultendpunct}{\mcitedefaultseppunct}\relax
\EndOfBibitem
\bibitem[Neder \emph{et~al.}(2017)Neder, Luxembourg, and Polman]{Neder2017}
V.~Neder, S.~L. Luxembourg and A.~Polman, \emph{Applied Physics Letters}, 2017,
  \textbf{111}, 073902\relax
\mciteBstWouldAddEndPuncttrue
\mciteSetBstMidEndSepPunct{\mcitedefaultmidpunct}
{\mcitedefaultendpunct}{\mcitedefaultseppunct}\relax
\EndOfBibitem
\bibitem[Flauraud \emph{et~al.}(2017)Flauraud, Reyes, Paniagua-Dominguez,
  Kuznetsov, and Brugger]{Flauraud2017}
V.~Flauraud, M.~Reyes, R.~Paniagua-Dominguez, A.~I. Kuznetsov and J.~Brugger,
  \emph{ACS Photonics}, 2017, \textbf{4}, 1913--1919\relax
\mciteBstWouldAddEndPuncttrue
\mciteSetBstMidEndSepPunct{\mcitedefaultmidpunct}
{\mcitedefaultendpunct}{\mcitedefaultseppunct}\relax
\EndOfBibitem
\bibitem[Chen \emph{et~al.}(2004)Chen, Taflove, and Backman]{Chen2004}
Z.~Chen, A.~Taflove and V.~Backman, \emph{Optics express}, 2004, \textbf{12},
  1214--1220\relax
\mciteBstWouldAddEndPuncttrue
\mciteSetBstMidEndSepPunct{\mcitedefaultmidpunct}
{\mcitedefaultendpunct}{\mcitedefaultseppunct}\relax
\EndOfBibitem
\bibitem[Chen \emph{et~al.}(2006)Chen, Taflove, and Backman]{Chen2006}
Z.~Chen, A.~Taflove and V.~Backman, \emph{Optics Letters}, 2006, \textbf{31},
  389--391\relax
\mciteBstWouldAddEndPuncttrue
\mciteSetBstMidEndSepPunct{\mcitedefaultmidpunct}
{\mcitedefaultendpunct}{\mcitedefaultseppunct}\relax
\EndOfBibitem
\bibitem[Ding \emph{et~al.}(2005)Ding, Lee, and Wang]{Ding2005}
L.~Ding, T.~Lee and C.-H. Wang, \emph{Journal of Controlled Release}, 2005,
  \textbf{102}, 395--413\relax
\mciteBstWouldAddEndPuncttrue
\mciteSetBstMidEndSepPunct{\mcitedefaultmidpunct}
{\mcitedefaultendpunct}{\mcitedefaultseppunct}\relax
\EndOfBibitem
\bibitem[Hong \emph{et~al.}(2008)Hong, Li, Yin, Li, and Zou]{Hong2008}
Y.~Hong, Y.~Li, Y.~Yin, D.~Li and G.~Zou, \emph{Journal of Aerosol Science},
  2008, \textbf{39}, 525--536\relax
\mciteBstWouldAddEndPuncttrue
\mciteSetBstMidEndSepPunct{\mcitedefaultmidpunct}
{\mcitedefaultendpunct}{\mcitedefaultseppunct}\relax
\EndOfBibitem
\bibitem[Bock \emph{et~al.}(2011)Bock, Woodruff, Hutmacher, and
  Dargaville]{Bock2011}
N.~Bock, M.~A. Woodruff, D.~W. Hutmacher and T.~R. Dargaville, \emph{Polymers},
  2011, \textbf{3}, 131--149\relax
\mciteBstWouldAddEndPuncttrue
\mciteSetBstMidEndSepPunct{\mcitedefaultmidpunct}
{\mcitedefaultendpunct}{\mcitedefaultseppunct}\relax
\EndOfBibitem
\bibitem[Bock \emph{et~al.}(2012)Bock, Dargaville, and Woodruff]{Bock2012}
N.~Bock, T.~R. Dargaville and M.~A. Woodruff, \emph{Progress in polymer
  science}, 2012, \textbf{37}, 1510--1551\relax
\mciteBstWouldAddEndPuncttrue
\mciteSetBstMidEndSepPunct{\mcitedefaultmidpunct}
{\mcitedefaultendpunct}{\mcitedefaultseppunct}\relax
\EndOfBibitem
\bibitem[Keshmiri and Troczynski(2002)]{KES2002}
M.~Keshmiri and T.~Troczynski, \emph{Journal of non-crystalline solids}, 2002,
  \textbf{311}, 89--92\relax
\mciteBstWouldAddEndPuncttrue
\mciteSetBstMidEndSepPunct{\mcitedefaultmidpunct}
{\mcitedefaultendpunct}{\mcitedefaultseppunct}\relax
\EndOfBibitem
\bibitem[Li \emph{et~al.}(2010)Li, Shen, Zong, and Yang]{LI2010}
S.~Li, Q.~Shen, J.~Zong and H.~Yang, \emph{Journal of Alloys and Compounds},
  2010, \textbf{508}, 99--105\relax
\mciteBstWouldAddEndPuncttrue
\mciteSetBstMidEndSepPunct{\mcitedefaultmidpunct}
{\mcitedefaultendpunct}{\mcitedefaultseppunct}\relax
\EndOfBibitem
\bibitem[Shiba and Ogawa(2009)]{SHI2009}
K.~Shiba and M.~Ogawa, \emph{Chemical Communications}, 2009,  6851--6853\relax
\mciteBstWouldAddEndPuncttrue
\mciteSetBstMidEndSepPunct{\mcitedefaultmidpunct}
{\mcitedefaultendpunct}{\mcitedefaultseppunct}\relax
\EndOfBibitem
\bibitem[Liu \emph{et~al.}(2017)Liu, Zhou, Zhang, Li, Chai, Fan, and
  Shao]{LIU2017}
G.~Liu, L.~Zhou, G.~Zhang, Y.~Li, L.~Chai, Q.~Fan and J.~Shao, \emph{Materials
  \& Design}, 2017, \textbf{114}, 10--17\relax
\mciteBstWouldAddEndPuncttrue
\mciteSetBstMidEndSepPunct{\mcitedefaultmidpunct}
{\mcitedefaultendpunct}{\mcitedefaultseppunct}\relax
\EndOfBibitem
\bibitem[Simeone \emph{et~al.}(2013)Simeone, Baldinozzi, Gosset, Le~Caer, and
  B{\'e}rar]{Simeone2013}
D.~Simeone, G.~Baldinozzi, D.~Gosset, S.~Le~Caer and J.-F. B{\'e}rar,
  \emph{Thin Solid Films}, 2013, \textbf{530}, 9--13\relax
\mciteBstWouldAddEndPuncttrue
\mciteSetBstMidEndSepPunct{\mcitedefaultmidpunct}
{\mcitedefaultendpunct}{\mcitedefaultseppunct}\relax
\EndOfBibitem
\bibitem[http://www.philiplaven.com/mieplot.htm()]{Mie_code}
http://www.philiplaven.com/mieplot.htm\relax
\mciteBstWouldAddEndPuncttrue
\mciteSetBstMidEndSepPunct{\mcitedefaultmidpunct}
{\mcitedefaultendpunct}{\mcitedefaultseppunct}\relax
\EndOfBibitem
\end{mcitethebibliography}
\bibliographystyle{rsc} 


%
\clearpage
\clearpage
\clearpage
\clearpage
\clearpage
\clearpage
\clearpage
\begin{figure}[t!]
\centering
\includegraphics[width=0.7\textwidth]{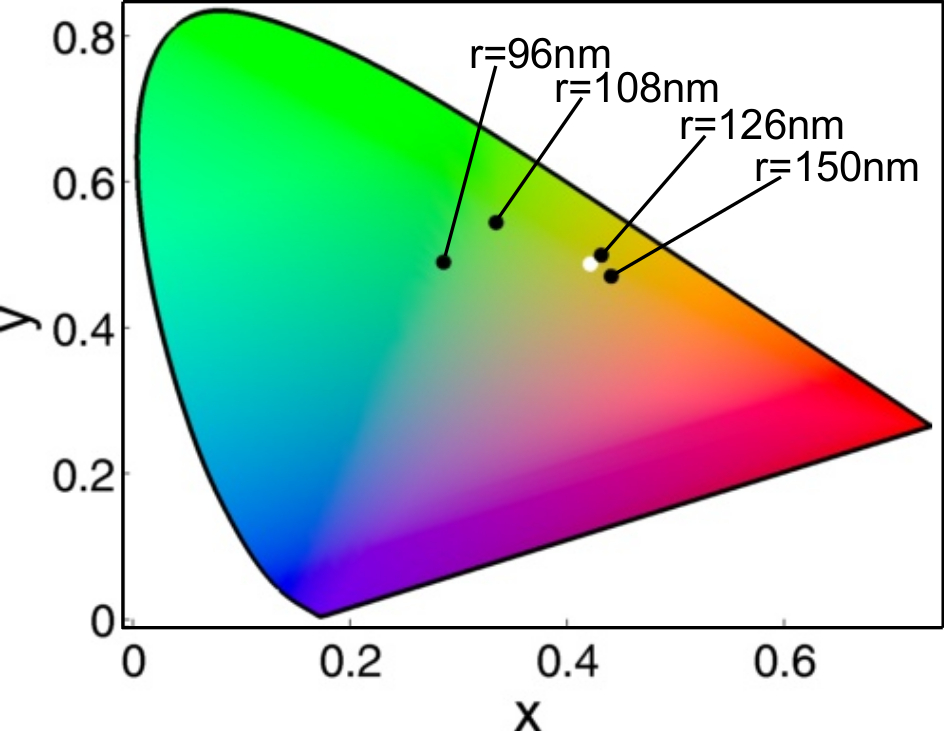}
\caption{CIE chromaticity gamut for light scattering in dark-field configuration (as those shown in Fig.\ref*{fgr:exp} (c)) for selected TiO$_2$ spheres with radius ranging from 96 to 150 nm. In these cases the spectra are non-normalized). The position of the light source used for illumination is specified in the map as a white full circle.}
\label{fgr:gamut}
\end{figure}
\clearpage

\end{document}